\documentclass{nature}
\usepackage{graphicx}
\usepackage{dcolumn}
\usepackage{bm}
\usepackage{color}
\usepackage{lineno}
\usepackage{epsfig}

\title{Observation of surface superconductivity in a three-dimensional Dirac material}

\author{Qi Liu,$^{1,\dag}$ Peng-Jie Guo,$^{2,\dag}$ Xiao-Yu Yue,$^{1}$ Zhe-Kai Yi,$^{1}$ Qing-Xin Dong,$^{5}$ Hui Liang,$^{1}$ Dan-Dan Wu,$^{1}$ Yan Sun,$^{1}$ Qiu-Ju Li,$^{4}$ Wen-Liang Zhu,$^{6}$ Tian-Long Xia,$^{2,\ddag}$ Xue-Feng Sun$^{3,1,7,\S}$ and Yi-Yan Wang$^{1,*}$}

\begin{document}
\bibliographystyle{naturemag}
	\maketitle	
	\begin{affiliations}
		\item Institute of Physical Science and Information Technology, Anhui University, Hefei, Anhui 230601, China
        \item Department of Physics and Beijing Key Laboratory of Opto-electronic Functional Materials \& Micro-nano Devices, Renmin University of China, Beijing 100872, China
        \item Department of Physics and Key Laboratory of Strongly-Coupled Quantum Matter Physics (CAS), University of Science and Technology of China, Hefei, Anhui 230026, China
        \item School of Physics \& Materials Science, Anhui University, Hefei, Anhui 230601, China        
        \item Institute of Physics and Beijing National Laboratory for Condensed Matter Physics, Chinese Academy of Sciences, Beijing 100190, China
        \item School of Physics and Information Technology, Shaanxi Normal University, Xi'an 710062, China
        \item Collaborative Innovation Center of Advanced Microstructures, Nanjing University, Nanjing, Jiangsu 210093, China
	\end{affiliations}
	
	\leftline{$^\dag$These authors contributed equally to this work.}
	\leftline{$^\ddag$Corresponding authors: tlxia@ruc.edu.cn}
    \leftline{$^\S$Corresponding authors: xfsun@ustc.edu.cn}
    \leftline{$^*$Corresponding authors: wyy@ahu.edu.cn}\vspace*{1cm}
	

\begin{abstract}
Superconductivity becomes more interesting when it encounters dimensional constraint or topology, because it is of importance for exploring exotic quantum phenomena or developing superconducting electronics. Here we report the coexistence of naturally formed surface superconducting state and three-dimensional topological Dirac state in single crystals of BaMg$_2$Bi$_2$. The electronic structure obtained from the first-principles calculations demonstrates that BaMg$_2$Bi$_2$ is an ideal Dirac material, in which the Dirac point is very close to the Fermi level and no other energy band crosses the Fermi level. Superconductivity up to 4.77 K can be observed under ambient pressure in the measurements of resistivity. The angle dependent magnetoresistance reveals the two-dimensional characteristic of superconductivity, indicating that superconductivity occurs on the surface of the sample and is absent in the bulk state. Our study not only provides BaMg$_2$Bi$_2$ as a suitable platform to study the interplay between superconductivity and topological Dirac state, but also indicates that MgBi-based materials may be a promising system for exploring new superconductors.
\end{abstract}

The study of two-dimensional (2D) superconductivity provides insight into a variety of quantum phenomena in condensed matter physics and material science\cite{saito2016highly}. The 2D superconductivity can be realized in systems such as heterogeneous interfaces or atomically ultrathin materials\cite{saito2016highly}. The 2D interfacial superconductivity has been observed in LaAlO$_3$/SrTiO$_3$\cite{reyren2007superconducting}, FeSe/SrTiO$_3$\cite{wang2012interface,ge2015superconductivity} and EuO/KTaO$_3$(111)\cite{liu2021two} \emph{etc}. Few-layer stanene and 1\emph{T}$_d$-MoTe$_2$ also exhibit superconductivity at low temperature\cite{liao2018superconductivity,cui2019transport}. In addition, superconductivity up to 14.9 K was discovered in thin film of infinite-layer nickelate Nd$_{0.8}$Sr$_{0.2}$NiO$_2$\cite{li2019superconductivity}, indicating the possibility of a family of nickelate superconductors. Recently, surface superconductivity can also be found in some topological materials\cite{shen2020two}, such as Dirac semimetal KZnBi\cite{PhysRevX.11.021065}, Cd$_3$As$_2$ nanoplate (proximity-induced)\cite{huang2019proximity} and type II Weyl semimetal TaIrTe$_4$\cite{xing2020surface}, which provide more platforms for studying 2D superconductivity.

Searching for superconductivity in topological materials has attracted a lot of interest for the possibility of obtaining novel topological phases, for example, the topological superconductivity (TSC)\cite{PhysRevLett.123.027003,PhysRevLett.115.187001}. Topological superconductors host superconducting bulk state and gapless Majorana surface state\cite{qi2011topological,sato2017topological}. The Majorana fermions exist in the superconducting gap and have the potential to be applied to topological quantum computation. Recently, Zhang \emph{et al.} reported the observation of TSC on the surface states of FeTe$_{0.55}$Se$_{0.45}$\cite{zhang2018observation}. Liu \emph{et al.} demonstrated the Majorana zero mode arising from the chiral topological surface state of (Li$_{0.84}$Fe$_{0.16}$)OHFeSe\cite{PhysRevX.8.041056}. The TSC can also be induced by the proximity effect between an \emph{s}-wave superconductor and surface Dirac fermions of topological insulator\cite{PhysRevLett.100.096407}. Experimentally, many attempts have been made to induce superconductivity on the surface of topological materials, including the fabrication of heterostructure\cite{wang2012coexistence,PhysRevLett.114.017001,PhysRevLett.112.217001,zhu2020interfacial,zhu2020superconducting} and tip/proximity-induced superconductivity\cite{huang2019proximity,wang2017discovery,wang2016observation,aggarwal2016unconventional} \emph{etc}.

In this work we grow the single crystals of BaMg$_2$Bi$_2$ and study the transport properties and electronic structure. The first-principles calculations show that BaMg$_2$Bi$_2$ is a three-dimensional (3D) Dirac material, in which the Dirac point exists in the direction of $\Gamma$-A and is protected by the symmetry. The Dirac point is very close to the Fermi level. In addition, only a small hole-type Fermi surface carrying the Dirac fermions crosses the Fermi level in the band structure, indicating that BaMg$_2$Bi$_2$ is an ideal material to explore extraordinary properties associated with Dirac fermions. Interestingly, superconductivity emerges below 4.77 K in the measurements of resistivity. The polar plots of the angle dependent magnetoresistance (MR) show a two-fold symmetry in the superconducting state, suggesting the 2D characteristic of the observed superconductivity. Our findings demonstrate the coexistence of surface superconductivity and 3D topological Dirac state in BaMg$_2$Bi$_2$.

Figure 1a illustrates the crystal structure of BaMg$_2$Bi$_2$, which consists of alternating Ba atoms and MgBi layers. BaMg$_2$Bi$_2$ crystalizes in the hexagonal CaAl$_2$Si$_2$-type structure with the space group of $P\bar{3}m1$ (No. 164). The crystal structure has C$_3$ rotational symmetry, as shown in the view from \emph{c}-axis direction (Fig. 1b). The first-principles calculations have been employed to study the electronic structure of BaMg$_2$Bi$_2$. When the spin-orbital coupling (SOC) effect is ignored, the band structure exhibits a gap around the $\Gamma$ point (Fig. S7 in the Supplementary Information). As shown in Figs. 1c and 1d, when the SOC effect is included, a Dirac point appears in the $\Gamma$-A direction, which is protected by the C$_3$ symmetry. It should be noted that the Dirac point is very close to the Fermi level and locates only about 8.5 meV above the Fermi level. Except for the Dirac fermion, there is no other energy band near the Fermi level. From the band structure, Dirac fermion will dominate the transport properties of BaMg$_2$Bi$_2$. The calculated band structure is consistent with the previous report on angle-resolved photoemission spectroscopy (ARPES) experiments\cite{takane2021dirac}, revealing the 3D topological Dirac state in BaMg$_2$Bi$_2$.

Figure 2a presents the temperature dependent resistivity $\rho$(\emph{T}) of BaMg$_2$Bi$_2$. The applied current is 100 $\mu$A. At first, the resistivity decreases with the decrease of temperature, exhibiting the metallic behavior. After the minimum at 160 K, an upturn appears in the resistivity curve with decreasing temperature, similar to a metal-insulator transition. Surprisingly, a sharp resistivity drop shows up at low temperature, indicating the transition to superconducting state. The superconducting transition temperature determined by linear epitaxy of normal curve and falling curve is 4.77 K (inset in Fig. 2a). The observed superconducting state has been further characterized by changing magnetic field and current intensity. Figure 2b shows the temperature dependent resistivity of BaMg$_2$Bi$_2$ under various magnetic fields. The applied field is perpendicular to the current direction. Superconductivity is gradually suppressed with increasing the field. At 8 T and 9 T, there is only a small drop in resistivity. The temperature dependent resistivity with different currents is shown in Fig. 2c. The superconducting transition temperature shifts toward lower temperatures with increasing current. When the current reaches 4 mA, the superconductivity is almost completely suppressed.

Figures 3a and 3b display the temperature dependence of critical field \emph{H}$_{c2}$ and critical current \emph{I}$_c$ extracted from Figs. 2b and 2c, respectively. As shown by the fit (red solid line) in Fig. 3a, the variation of critical field can be well described by the empirical formula \emph{H}$_{c2}$ = \emph{H}$_{c2}$(0)[1 -- (\emph{T}/\emph{T}$_c$)$^2$]. The obtained upper critical field at 0 K is 8.7 T. With the increase of current, \emph{T}$_c$ decreases rapidly. The behavior is similar to the interfacial superconductivity in TaAs/Ag heterostructure and LaAlO$_3$/SrTiO$_3$ system.

Since BaMg$_2$Bi$_2$ exhibits a superconducting transition in the measurement of resistivity, the Meissner effect and specific heat jump are expected. However, after testing on several samples, it turns out that although superconducting transition can be observed in resistivity, the Meissner effect and a jump in heat capacity are absent (Figs. S2 and S3 in the Supplementary Information). This implies the possibility of surface superconductivity in the samples of BaMg$_2$Bi$_2$. Superconductivity only occurs on the surface of the sample, but the bulk state accounts for a large proportion in the measurements of magnetic susceptibility and specific heat.

In order to further understand the unusual superconductivity in BaMg$_2$Bi$_2$, the measurements of angle dependent resistivity under magnetic field are performed. The schematic diagram of the measurements are plotted in Figs. 4a and 4d. The current is along the hexagonal edge of crystal and the magnetic field rotates in the \emph{yz}-plane and \emph{xz}-plane, respectively. In the former configuration, as shown in Figs. 4b and 4c, the polar plots of the angle dependent MR exhibit two-fold symmetry with a period of $\pi$. The maximum and minimum values of resistivity appear at 0$^{\circ}$, 180$^{\circ}$ (\emph{B} $\bot$ \emph{ab} plane) and 90$^{\circ}$, 270$^{\circ}$ (\emph{B} // \emph{ab} plane), respectively. In the latter configuration, the patterns in Figs. 4e and 4f show a slight distortion, but the two-fold symmetry remains. It should be noted that the two-fold symmetry only exists in the superconducting state. Once the superconducting state is gradually suppressed with the increase of magnetic field or temperature, the two-fold symmetry will be weakened and disappear. In the nonsuperconducting state, the angle dependent MR shows isotropic characteristic, which comes from the 3D Fermi surface of the bulk state. In general, the two-fold symmetry means that the conductive layer is 2D. In materials, the Lorentz force affects the carrier momentum component in the plane perpendicular to the magnetic field. For a 2D conductive layer, the carrier only responds to the magnetic field component \emph{B}$\mid$cos$\theta$$\mid$, resulting in the two-fold symmetry of the polar plots. In addition, the observed two-fold symmetry cannot be derived from the bulk state, because the bulk state is a small 3D Fermi surface, as demonstrated by the first-principles calculations and previous ARPES experiment. Thus, the superconductivity in BaMg$_2$Bi$_2$ is considered to be a 2D superconductivity that occurs on the surface of the sample.

In fact, the properties of BaMg$_2$Bi$_2$ are similar to KZnBi, which also hosts the surface superconductivity (0.85 K) and Dirac state\cite{PhysRevX.11.021065}. In both samples, the superconductivity naturally forms on the surface without external stimuli while the bulk state remains nonsuperconducting. Since Dirac fermions exist in the bulk state, it is worth exploring the relationship or interplay between the superconductivity and topological state. Further researches (for example, ARPES experiment in superconducting state) are needed to clarify the remaining problems. On the material side, BaMg$_2$Bi$_2$ possesses the MgBi layers that are important for transport properties and the formation of Dirac state. The MgBi-based materials may provide more opportunities for exploring new superconductors.

In summary, we report the observation of surface superconductivity up to 4.77 K under ambient pressure in the single crystals of BaMg$_2$Bi$_2$, which hosts the 3D topological Dirac state. The Dirac state has been described by the band structure from first-principles calculations. The superconductivity was characterized by the measurements of resistivity. The 2D nature of superconductivity was verified by the angle dependent MR, supporting that the superconductivity in BaMg$_2$Bi$_2$ occurs on the surface rather than in the bulk.

\section*{Methods}
\subsection{Crystal growth and transport measurements}
Large single crystals of BaMg$_2$Bi$_2$ were grown by flux method. The starting elements, Ba, Mg and Bi, were mixed and put into the alumina crucible with a molar ratio of Ba : Mg : Bi=1 : 5 : 9. The crucible was then sealed into an evacuated quartz tube and heated to 900$^\circ$C. After cooling to 650$^{\circ}$C in 300 h, the excess flux was removed with a centrifuge. The atomic composition of BaMg$_2$Bi$_2$ has been checked by energy dispersive x-ray spectroscopy. The measurements of transport properties were performed on a Quantum Design physical property measurement system (PPMS). The obtained crystals are sensitive to both air and GE varnish. The samples can survive in air for only several minutes. The electrodes used in the measurements were made in a glove box and then covered with a layer of N-grease.

\subsection{Electronic structure calculations}
The first-principles electronic structure calculations were performed with the projector augmented wave (PAW) method\cite{PhysRevB.50.17953,PhysRevB.59.1758} as implemented in the Vienna \emph{ab initio} simulation package (VASP)\cite{PhysRevB.47.558,kresse1996efficiency,PhysRevB.54.11169}. The generalized gradient approximation (GGA) of Perdew-Burke-Ernerh (PBE) type\cite{PhysRevLett.77.3865} was adopted for the exchange-correlation potential. The kinetic energy cutoff of the plane wave basis was set to be 400 eV. For the Brillouin zone sampling, a 15$\times$15$\times$9 \emph{k}-point mesh was employed. The Gaussian smearing with a width of 0.02 eV was used around the Fermi surface. In structural optimization, both cell parameters and internal atomic positions were allowed to relax until all forces on atoms were smaller than 0.01 eV/{\AA}. The relaxed structure parameters are well agreed with experimental results. The spin-orbital-coupling (SOC) effect was included in the calculations of the electronic properties.

\section*{Data availability}
The authors declare that the data supporting the findings of this study are available within the article and its Supplementary Information. Extra data are available from the corresponding authors upon reasonable request.

\section*{References}
\bibliography{Bibtex}

\begin{addendum}
\item This work is supported by the National Natural Science Foundation of China (Nos.12104011, 11874336, 12074425, 11874422, 11904003, 12104010, 52102333, 12004003), the Natural Science Foundation of Anhui Province (Nos.2108085QA22, 1908085MA09, 2108085MA16), the Joint Funds of the National Natural Science Foundation of China (No. U1832209) and the National Key R$\&$D Program of China (No. 2019YFA0308602).

\item[Author contributions] Y.-Y.W. coordinated the project and designed the experiments. Q.L. and Y.-Y.W. synthesized the single crystals of BaMg$_2$Bi$_2$. P.-J.G. performed \emph{ab initio} calculations. Y.-Y.W. performed the transport measurements with the assistance of Q.L. and X.-Y.Y.. Q.L., Y.-Y.W. and P.-J.G. plotted the figures and analysed the experimental data. Y.-Y.W., X.-F.S. and T.-L.X. wrote the paper. All authors discussed the results and commented on the manuscript.

\item[Supplementary Information] accompanies this paper.

\item[Author Information] The authors declare no competing interests. The data that support the findings of this study are available from the corresponding authors T.-L.X. (tlxia@ruc.edu.cn), X.-F.S. (xfsun@ustc.edu.cn) and Y.-Y.W. (wyy@ahu.edu.cn) upon reasonable request.

\end{addendum}
\newpage

\begin{figure*}
	\centerline{\epsfig{figure=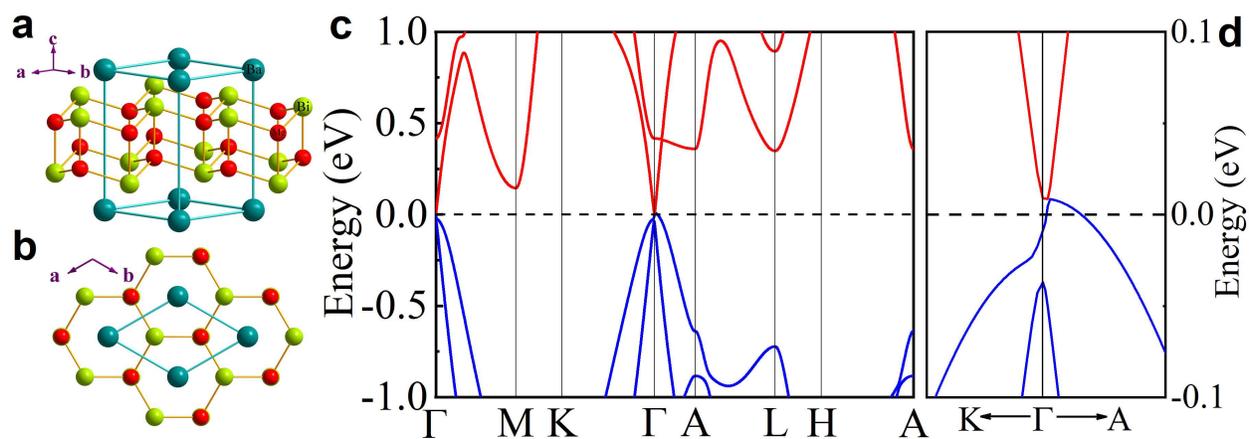,width=\columnwidth}}
	\caption{\textbf{Crystal structure and electronic structure of BaMg$_2$Bi$_2$.} \textbf{a}, Schematic crystal structure of BaMg$_2$Bi$_2$. \textbf{b},View of the structure from the direction of \emph{c}-axis. \textbf{c}, The calculated band structure of BaMg$_2$Bi$_2$ along high symmetry lines with the SOC effect included. \textbf{d}, The enlarged view of band structure around $\Gamma$ point.}
\end{figure*}

\begin{figure*}
	\centerline{\epsfig{figure=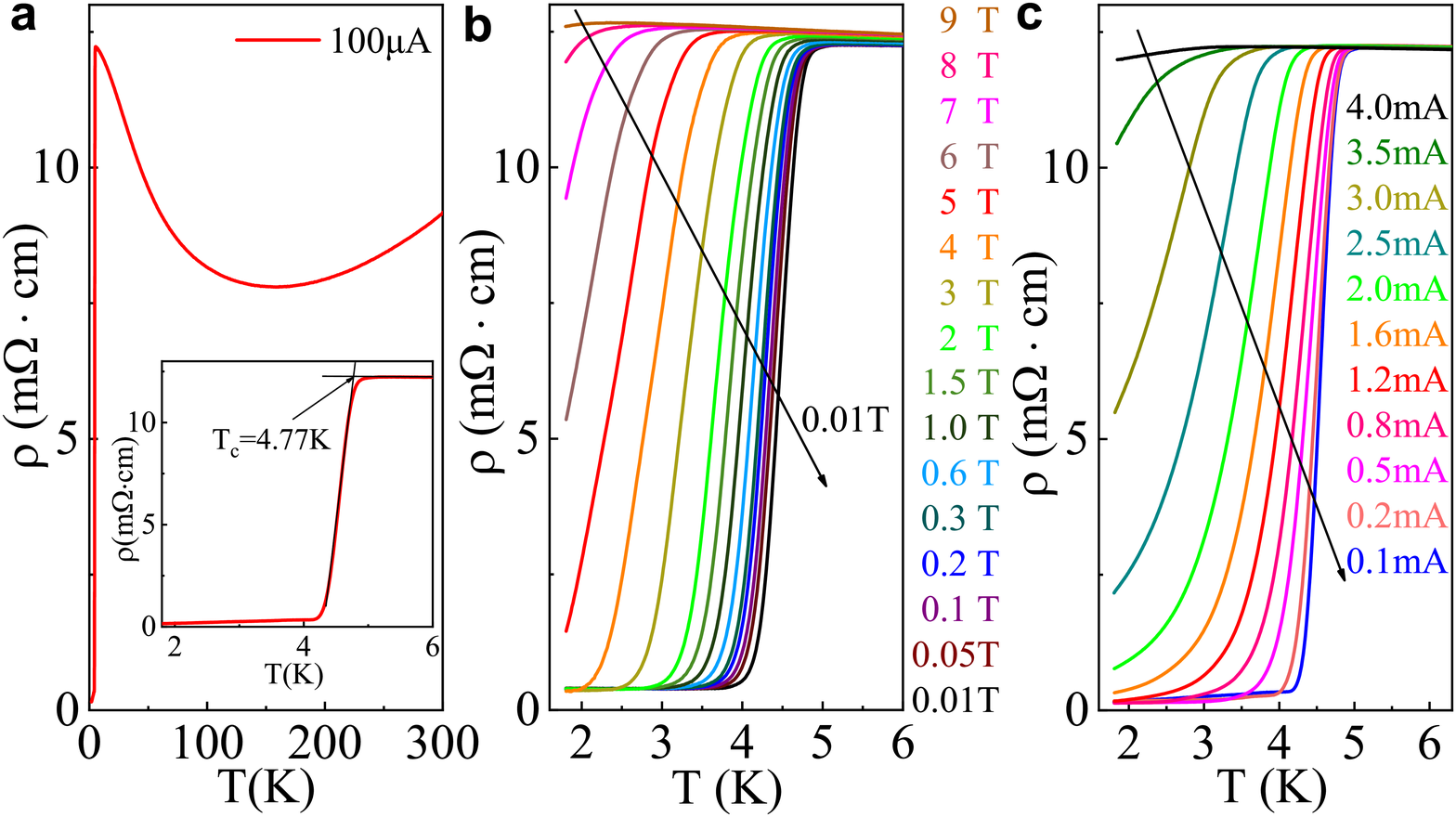,width=\columnwidth}}
	\caption{\textbf{Superconductivity of the BaMg$_2$Bi$_2$ single crystals.} \textbf{a}, Temperature dependence of zero field resistivity $\rho$(\emph{T}) in the range of 1.8 K to 300 K. Inset: the enlarged part from 1.8 K to 6 K. \textbf{b}, The $\rho$(\emph{T}) curves under different magnetic fields at low temperature. The applied current is 100 $\mu$A. \textbf{c}, The $\rho$(\emph{T}) curves under various currents at low temperature and zero magnetic field.}
\end{figure*}

\begin{figure*}
	\centerline{\epsfig{figure=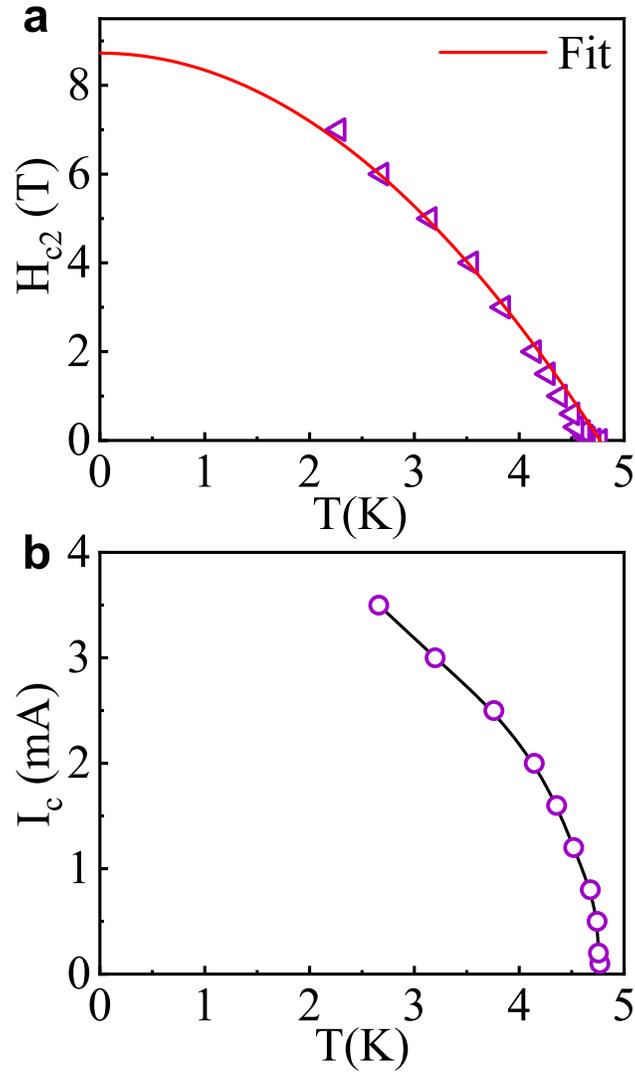,width=0.5\columnwidth}}
	\caption{\textbf{The critical field and critical current of the superconducting state.} \textbf{a}, Temperature dependence of critical field extracted from Fig. 2b. The red line is a fit to the data using the empirical formula. \textbf{b}, Temperature dependence of critical current obtained from Fig. 2c.}
\end{figure*}

\begin{figure*}
	\centerline{\epsfig{figure=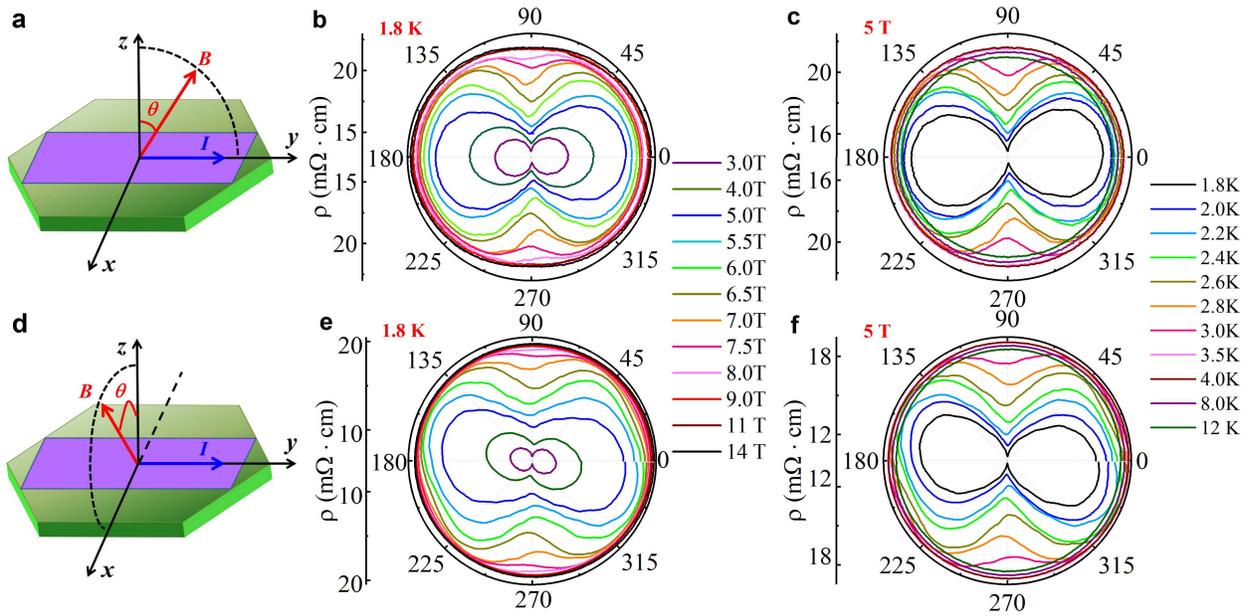,width=\columnwidth}}
	\caption{\textbf{Angle dependent magnetoresistance in the superconducting and nonsuperconducting states.} \textbf{a, d}, Schematic of two configurations of the measurements. The current is along the hexagonal edge and the magnetic field rotates in the \emph{yz}-plane (\textbf{a}) and \emph{xz}-plane (\textbf{d}), respectively. \textbf{b, e}, Polar plots of the angle dependent resistivity corresponding to the configurations in \textbf{a} and \textbf{d}. The temperature is fixed at 1.8 K and the magnetic field varies from 3 T to 14 T. \textbf{c, f}, Polar plots of the angle dependent resistivity corresponding to the configurations in \textbf{a} and \textbf{d}. The magnetic field is fixed at 5 T and the temperature changes from 1.8 K to 12 K.}
\end{figure*}

\end{document}


\title{Supplementary Information for: Observation of surface superconductivity in a three-dimensional Dirac material}


\author{Qi Liu,$^{1,\dag}$ Peng-Jie Guo,$^{2,\dag}$ Xiao-Yu Yue,$^{1}$ Zhe-Kai Yi,$^{1}$ Qing-Xin Dong,$^{5}$ Hui Liang,$^{1}$ Dan-Dan Wu,$^{1}$ Yan Sun,$^{1}$ Qiu-Ju Li,$^{4}$ Wen-Liang Zhu,$^{6}$ Tian-Long Xia,$^{2,\ddag}$ Xue-Feng Sun$^{3,1,7,\S}$ and Yi-Yan Wang$^{1,*}$}

	\maketitle
	
	\begin{affiliations}
		\item Institute of Physical Science and Information Technology, Anhui University, Hefei, Anhui 230601, China
        \item Department of Physics and Beijing Key Laboratory of Opto-electronic Functional Materials \& Micro-nano Devices, Renmin University of China, Beijing 100872, China
        \item Department of Physics and Key Laboratory of Strongly-Coupled Quantum Matter Physics (CAS), University of Science and Technology of China, Hefei, Anhui 230026, China
        \item School of Physics \& Materials Science, Anhui University, Hefei, Anhui 230601, China
        \item Institute of Physics and Beijing National Laboratory for Condensed Matter Physics, Chinese Academy of Sciences, Beijing 100190, China
        \item School of Physics and Information Technology, Shaanxi Normal University, Xi'an 710062, China
        \item Collaborative Innovation Center of Advanced Microstructures, Nanjing University, Nanjing, Jiangsu 210093, China
	\end{affiliations}
	
	\leftline{$^\dag$These authors contributed equally to this work.}
	\leftline{$^\ddag$Corresponding authors: tlxia@ruc.edu.cn}
    \leftline{$^\S$Corresponding authors: xfsun@ustc.edu.cn}
    \leftline{$^*$Corresponding authors: wyy@ahu.edu.cn}\vspace*{1cm}
	
\newpage

\begin{figure*}
	\centerline{\epsfig{figure=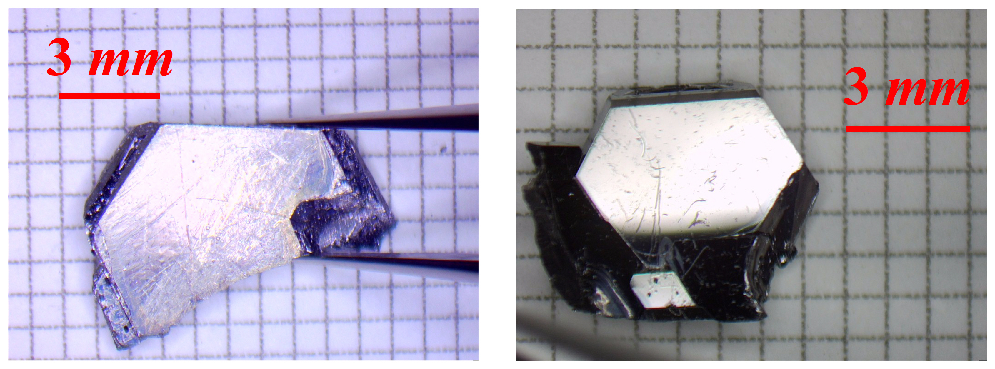,width=0.8\columnwidth}}
	\caption{Two typical images of the grown single crystals of BaMg$_2$Bi$_2$. The obtained crystals are as large as 9 mm in its one dimension. The hexagonal feature can be clearly seen from the pictures.}
\end{figure*}

\begin{figure*}
	\centerline{\epsfig{figure=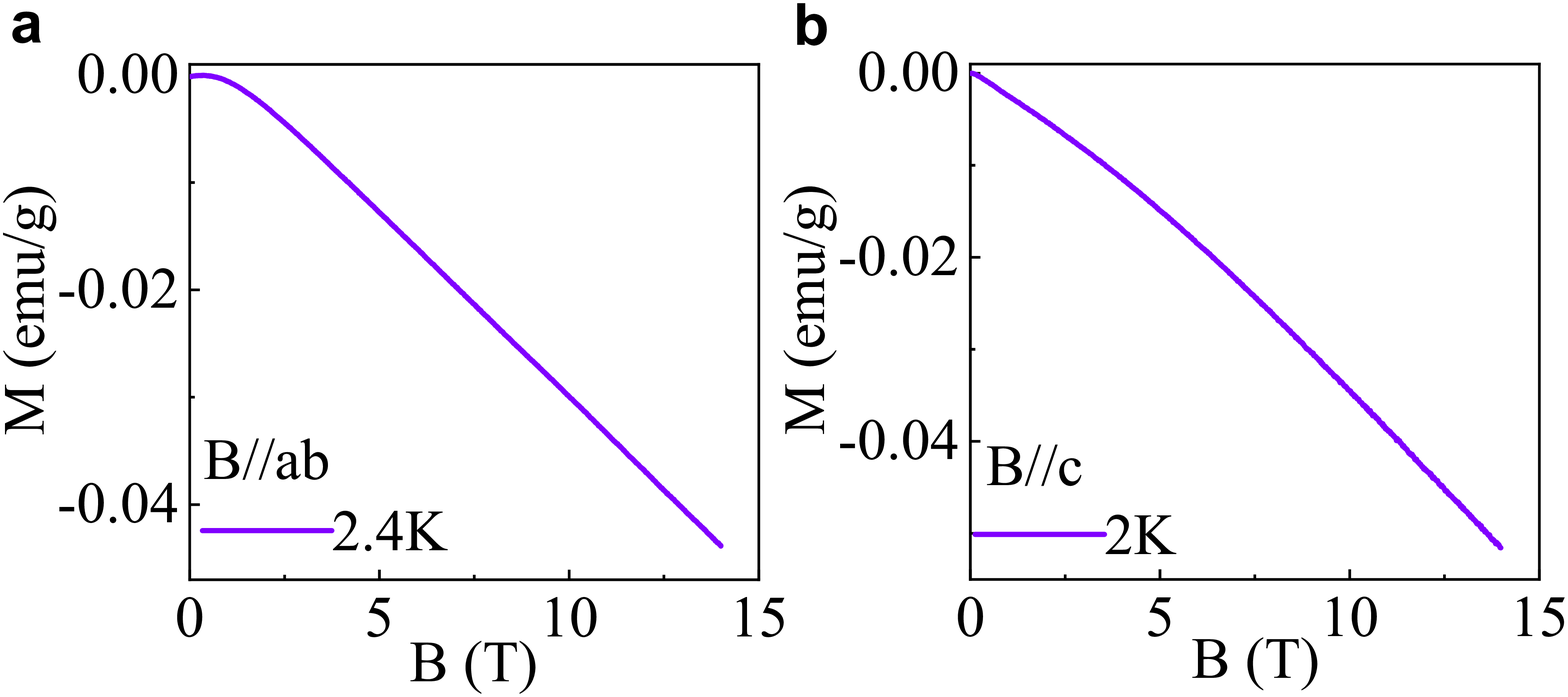,width=0.8\columnwidth}}
	\caption{Magnetization versus magnetic field for \emph{B}//\emph{ab} (\textbf{a}) and \emph{B}//\emph{c} (\textbf{b}). The Meissner effect is absent in the test of magnetic susceptibility.}
\end{figure*}

\begin{figure*}
	\centerline{\epsfig{figure=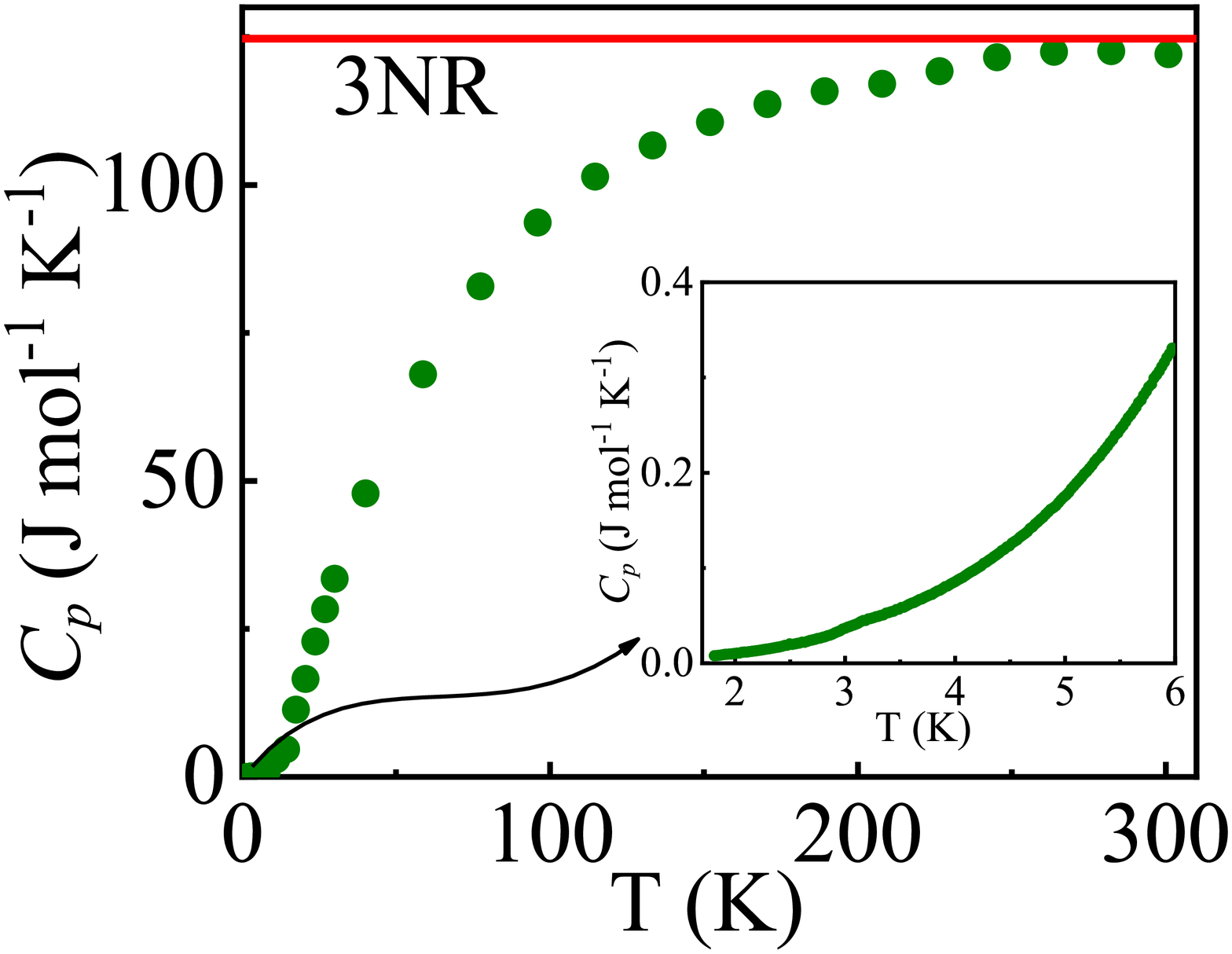,width=0.5\columnwidth}}
	\caption{The specific heat of BaMg$_2$Bi$_2$ from 1.8 K to 300 K. The solid horizontal line shows the classic value of Dulong-Petit law at high temperature limit. As shown in the inset, the specific heat jump cannot be observed at the superconducting transition temperature.}
\end{figure*}

\begin{figure*}
	\centerline{\epsfig{figure=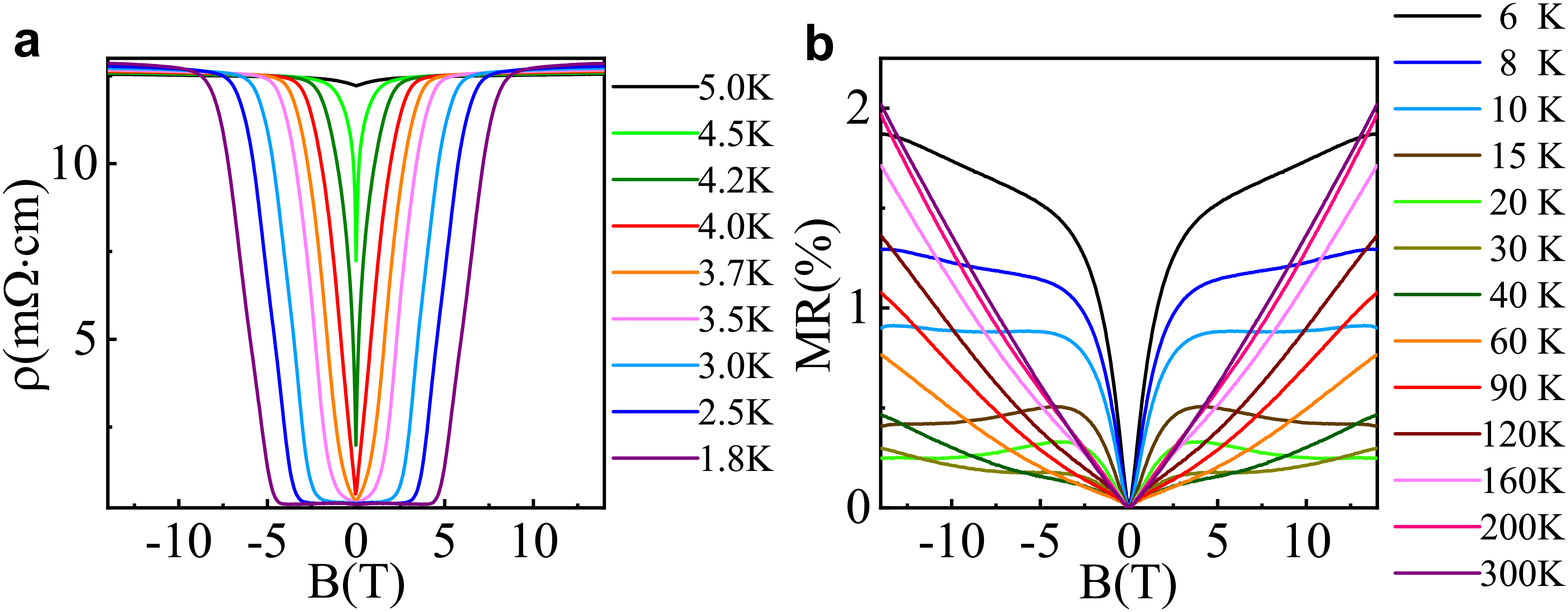,width=0.8\columnwidth}}
	\caption{\textbf{a}, Field dependent resistivity of superconducting states at various temperatures. \textbf{b}, Field dependent MR of nonsuperconducting states at various temperatures. The weak antilocalization effect can be found at low temperatures in nonsuperconducting states.}
\end{figure*}

\begin{figure*}
	\centerline{\epsfig{figure=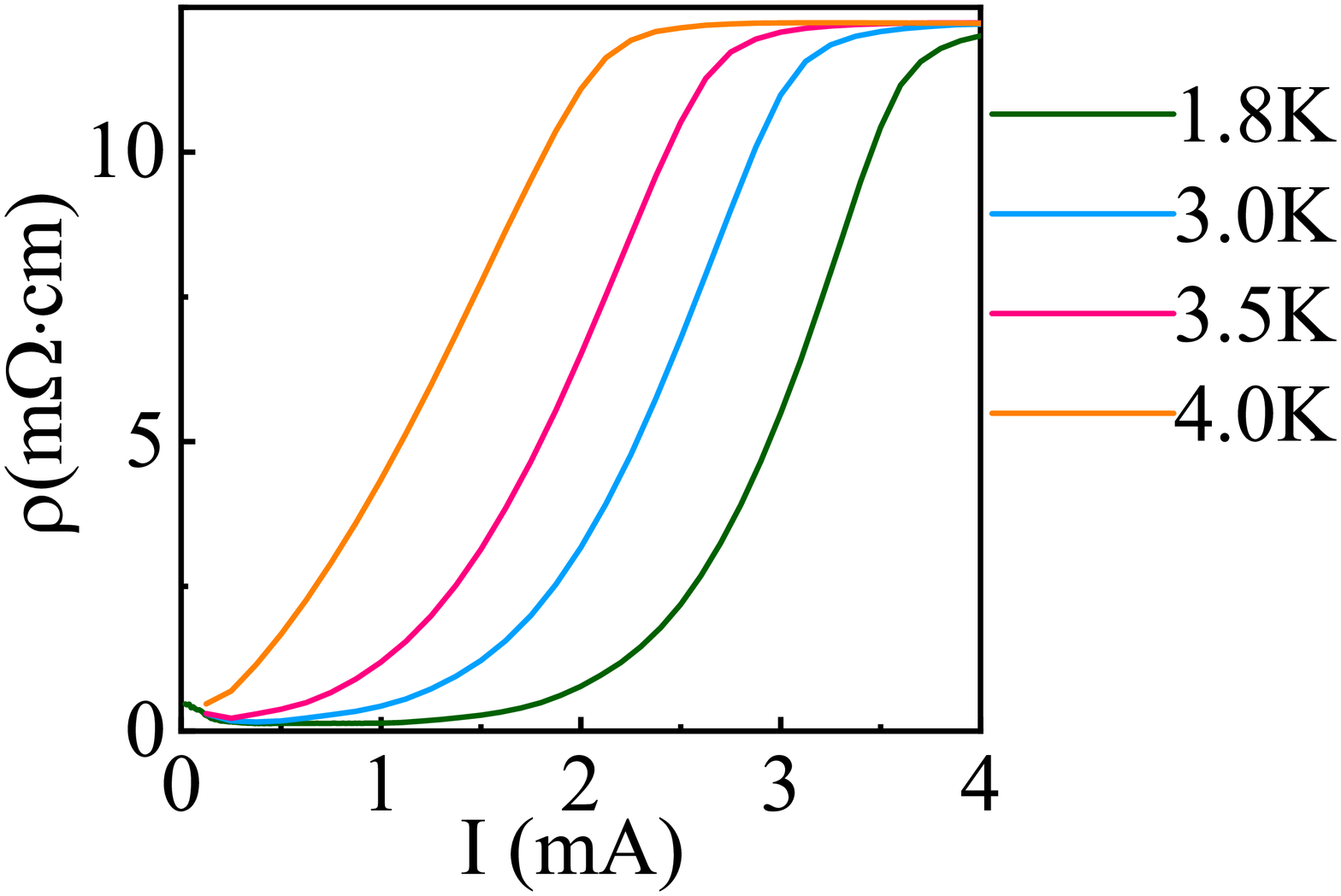,width=0.5\columnwidth}}
	\caption{Resistivity versus current in superconducting states.}
\end{figure*}

\begin{figure*}
	\centerline{\epsfig{figure=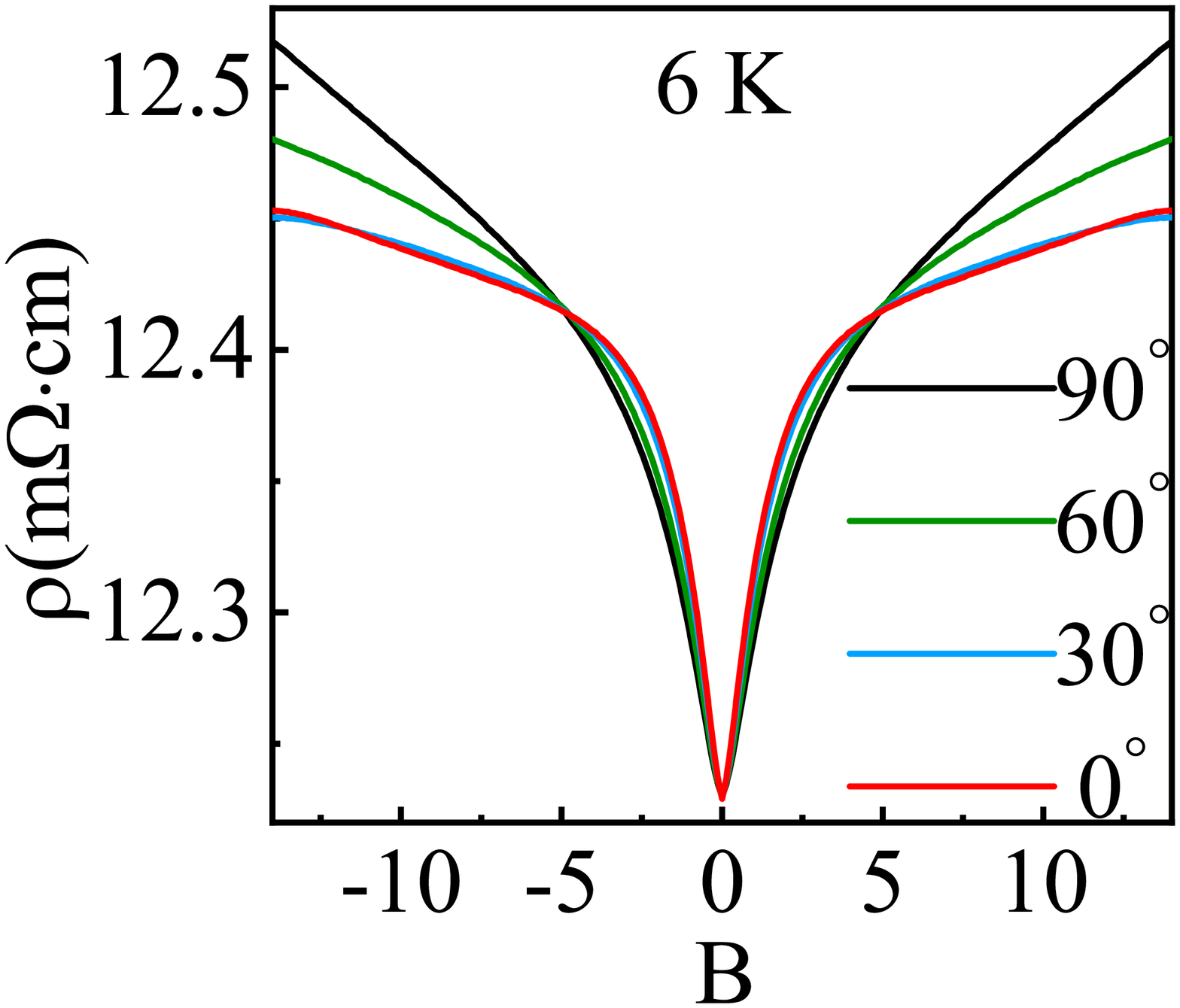,width=0.5\columnwidth}}
	\caption{Field dependent MR in nonsuperconducting states (6 K) at different angles. The corresponding configuration is the same as Fig. 4a in the text.}
\end{figure*}

\begin{figure*}
	\centerline{\epsfig{figure=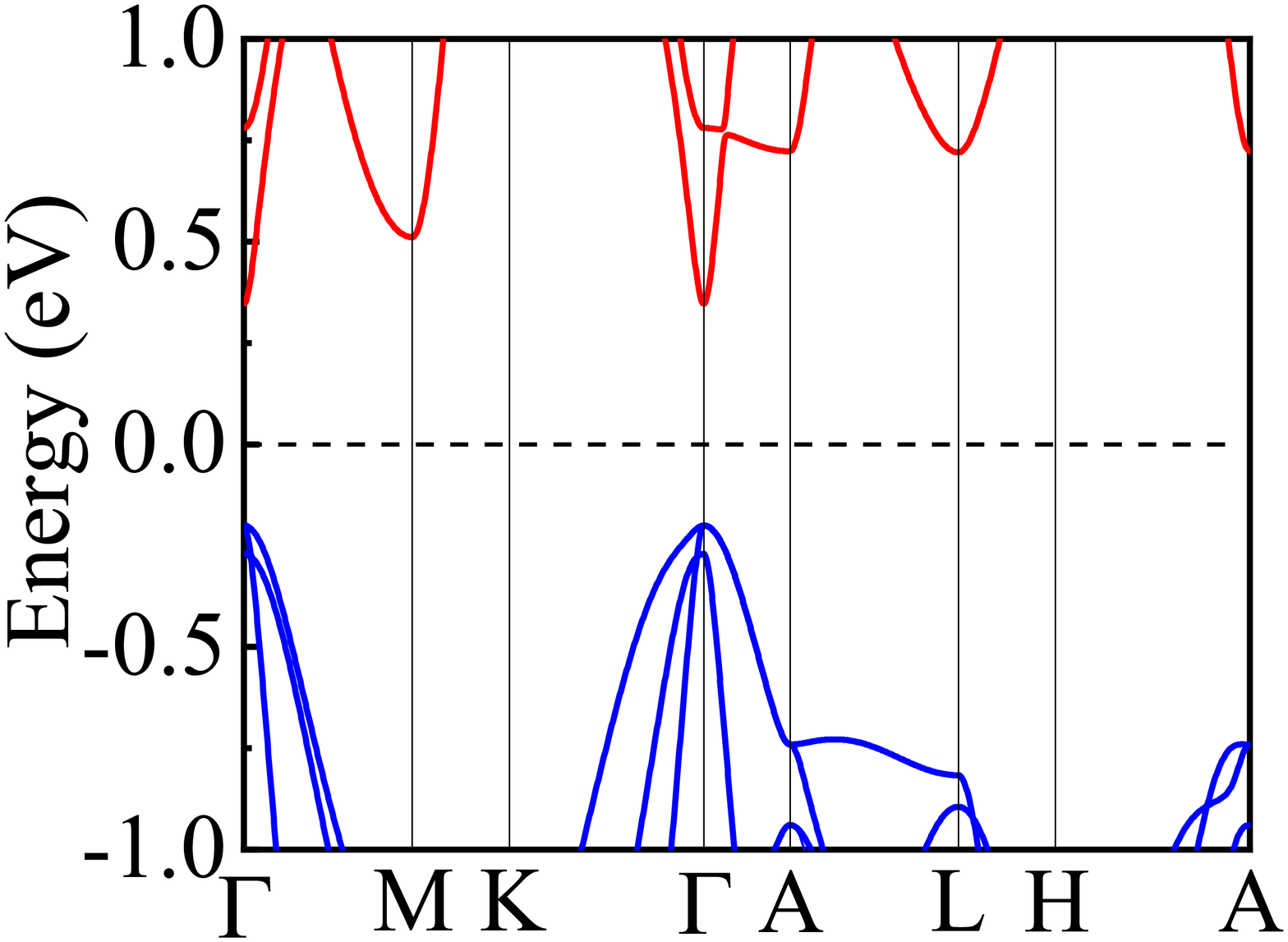,width=0.5\columnwidth}}
	\caption{The calculated band structure of BaMg$_2$Bi$_2$ along high symmetry lines in Brillouin zone with the SOC effect ignored.}
\end{figure*}